\documentclass[12pt]{article}
\usepackage[left=1in,top=1in,right=1in,bottom=1in,nohead]{geometry}
\usepackage{amsmath,amssymb,comment}
\usepackage{mathtools}
\usepackage{rotating}
\usepackage{graphicx}

\usepackage{subcaption}
\usepackage{float}
\usepackage{amsmath,amssymb}
\usepackage{amssymb}
\usepackage{booktabs}
\usepackage{amsfonts}
\usepackage{graphics,adjustbox}
\usepackage{bm}                
\usepackage{mathrsfs}      
\usepackage{amsmath, amsthm}
\usepackage{graphicx,epsf,pstricks,pst-node,psfrag,amsthm,amssymb,amsmath}
\usepackage{float}
\usepackage{array}
\usepackage{authblk}
\usepackage{natbib}
\usepackage{url}
\usepackage{hyperref}
\usepackage{rotating}
\usepackage{multicol}
\usepackage{multirow}

\usepackage{bm}

\newcommand{\wvec}{\ensuremath{\bm{w}}}

\newcommand{\vepsvec}{\ensuremath{{\bm{\varepsilon}}}}
\newcommand{\etavec}{\ensuremath{{\bm{\eta}}}}

\newcommand{\omegavec}{\ensuremath{\bm{\omega}}}

\newcommand{\muvec}{\ensuremath{{\bm{\mu}}}}

\newcommand{\thetavec}{\ensuremath{{\bm{\theta}}}}

\newcommand{\univec}{\bm{1}}
\DeclareMathOperator*{\argmin}{arg\,min}

\newcommand{\zerovec}{\ensuremath{{\bf 0}}}

\newcommand{\be}{\begin{equation}}
\newcommand{\ee}{\end{equation}}
\newcommand{\beqa}{\begin{eqnarray*}}
\newcommand{\eeqa}{\end{eqnarray*}}
\newcommand{\beqn}{\begin{eqnarray}}
\newcommand{\eeqn}{\end{eqnarray}}
\newcommand{\ba}{\begin{array}}
\newcommand{\ea}{\end{array}}
\newcommand{\bc}{\begin{center}}
\newcommand{\ec}{\end{center}}
\newcommand{\btab}{\begin{tabular}}
\newcommand{\etab}{\end{tabular}}

\newcommand{\mb}{\makebox}

\newcommand{\st}{\stackrel}

\newcommand{\ind}{\, \raise-2pt\hbox{$\st{\mb{\scriptsize ind}}{\sim}$}\, }
\newcommand{\iid}{\, \raise-2pt\hbox{$\st{\mb{\scriptsize iid}}{\sim}$}\,}

\newcommand{\bX}{\bm{X}}

\newcommand{\bGamma}{\bm{\Gamma}}

\newcommand{\bSigma}{\bm{\Sigma}}
\newcommand{\bOmega}{\bm{\Omega}}

\newcommand{\bmB}{\bm{B}}
\newcommand{\bmD}{\bm{D}}
\newcommand{\bmI}{\bm{I}}

\newcommand{\RR}{{\mathbb R}}

\mathchardef\given="626A
\mathcode`:="603A

\long\def\beginskip#1\endskip{}
\def\endskip{}

\begin{document}
	\title{High-dimensional Portfolio Optimization using Joint Shrinkage}
\date{} 
\author[1]{Anik Burman}
\author[2]{Sayantan Banerjee \footnote{Corresponding author. Address: Operations Management \& Quantitative Techniques Area, IIM Indore, Rau-Pithampur Road, Indore, MP 453556, India. e-mail: \url{sayantanb@iimidr.ac.in}}}

\affil[1]{Indian Statistical Institute, Kolkata}  
\affil[2]{Indian Institute of Management, Indore}
\maketitle
	
	\begin{abstract}
		We consider the problem of optimizing a portfolio of financial assets, where the number of assets can be much larger than the number of observations. The optimal portfolio weights require estimating the inverse covariance matrix of excess asset returns, classical solutions of which behave badly in high-dimensional scenarios. We propose to use a regression-based joint shrinkage method for estimating the partial correlation among the assets. Extensive simulation studies illustrate the superior performance of the proposed method with respect to variance, weight, and risk estimation errors compared with competing methods for both the global minimum variance portfolios and Markowitz mean-variance portfolios. We also demonstrate the excellent empirical performances of our method on daily and monthly returns of the components of the S\&P 500 index.
\end{abstract}
	
	\begin{quotation}
		\noindent {\it Keywords:} High-dimensional models; Joint shrinkage; Precision matrix; Portfolio optimization.
	\end{quotation}\par

\section{Introduction}

Portfolio optimization is a widely studied and relevant problem in portfolio management applications. Optimal portfolio selection methods with the global minimum variance approach or the Markowitz mean-variance model \citep{markowitz1952} are frequently used by portfolio managers, including robo-advisors \citep{sabharwal2018robo}. The optimal portfolio allocation weights and associated risk are functions of the mean and precision matrix (inverse covariance matrix) of the excess asset returns, thus making it imperative to estimate these parameters from the available data accurately. 

In most of modern era portfolio optimization problems, the number of assets $p$ is often large, exceeding the number of observations $n$. There might be multiple reasons for this. For example, if we consider a portfolio with $p = 500$ assets, even a 10-year long monthly returns data would still result in $p \gg n.$ The dynamics of interdependence among the various assets might also change over a long period, and for new assets, long history of returns is unavailable. All of these lead to a high-dimensional scenario, where classical statistical inference techniques fail to provide stable or robust solutions. Estimating a covariance matrix or the precision matrix is a non-trivial problem in such instances owing to singularity. 

Sparsity plays an important role in tackling the high-dimensional problem. This essentially refers to exploring the low-dimensional structure in a high dimensional space to make valid inference; see, for example, \cite{friedman2008sparse}. \cite{meinshausen2006} developed a regression-based estimator using lasso for estimating the entries of the precision matrix via regression coefficients. \cite{callot2021nodewise} proposed a similar nodewise regression approach for application in the portfolio optimization problem. However, their approach leads to an estimator that is not necessarily symmetric.

In this paper, we propose to use a regression-based joint shrinkage method developed by \cite{peng2009} to estimate the precision matrix of a portfolio of a large number of assets. The method is more efficient than competing approaches in terms of computational complexity and the number of parameters to be estimated. 

The paper is organized as follows. In the next section, we discuss the optimal solution for two different portfolio choices. We describe our proposed method in Section~\ref{sec:Method}. Numerical results are presented in Sections~\ref{sec:Simu} and \ref{sec:Real}, followed by a discussion. Details of simulations, additional real-data results, and visualizations are presented as supplementary materials.

\section{Optimum portfolios}
\label{optport}

Let us define $\bX^{(t)} = (X_{1}^{(t)},\ldots,X_{p}^{(t)})'$ as the random vector of excess asset returns at time $t \in \{1,\ldots,n\}$ with mean $\muvec$ and a $p$-dimensional positive definite covariance matrix $\bSigma$. The corresponding precision matrix is $\bOmega = \bSigma^{-1}$ with the $(i,j)$th element denoted by $\omega_{ij}.$ A portfolio comprising of $p$ assets is given by an asset allocation vector of weights $\wvec = (w_1,\ldots,w_p)' \in \RR^p$ such that $\wvec' \univec = 1.$ The global minimum variance portfolio is defined as
\begin{equation}
\label{eqn:GMVportfolio}
    \wvec_G = \argmin_{\wvec' \univec = 1} \wvec'\bSigma \wvec,
\end{equation}
where $\bSigma$ is the covariance matrix of excess returns of assets in the portfolio. The solution to the above optimization problem is 
$ \wvec_G = (\univec'\bSigma^{-1}\univec)^{-1}\bSigma^{-1}\univec,$
with the minimum expected risk of portfolio return given by $(\univec'\bSigma^{-1}\univec)^{-1}.$

The Markowitz mean-variance framework takes into account a fixed expected return, say $\mu^*$, and solves the optimization problem
\begin{equation}
\label{eqn:MPortfolio}
    \wvec_M = \argmin_{\wvec'\univec = 1, \wvec'\muvec = \mu^*} \wvec'\bSigma\wvec.
\end{equation}
The optimal weight for this problem is given by $\wvec_M = \{B\bOmega\univec - A\bOmega\muvec + \mu^*(C\bOmega\muvec - A\bOmega\univec)\}/D,$ where $A = \muvec'\bOmega\univec, B = \muvec'\bOmega\muvec, C = \univec'\bOmega\univec, D = BC - A^2.$

The estimated portfolio weights as given in expressions (\ref{eqn:GMVportfolio}) and (\ref{eqn:MPortfolio}) and their corresponding risks are obtained via plug-in estimates of $\muvec$ and $\bOmega$, respectively denoted by $\hat{\muvec}$ and $\hat{\bOmega}.$ 

\section{Methodology}
\label{sec:Method}
Our proposed methodology for portfolio optimization in the high-dimensional scenario depends on the intrinsic relationship between elements of the precision matrix and regression coefficients obtained via regressing each of the individual excess asset returns on the remainder of the same. We elaborate on this relationship in detail below.

\subsection{Regression approach to precision matrix estimation}

We consider $p$ separate linear regression models 
\[X_j = \sum_{j \neq k}\beta_{jk}X_k + \epsilon_j,\; j = 1,\ldots,p.\]
We can show that, for every $j = 1,\ldots,p,$ a necessary and sufficient condition for the error term $\epsilon_j$ and the set of predictors $X_{-j} := \{X_k: k \neq j\}$ to be uncorrelated is $\beta_{jk} = -\omega_{jk}/\omega_{jj}$. In addition to this, we also have $\mathrm{Var}(\epsilon_j) = 1/\omega_{jj}$ for every $j = 1,\ldots,p.$ Also, for $1 \leq i \neq j \leq p,$ the correlation between $\epsilon_i$ and $\epsilon_j$ is defined as the partial correlation $\rho^{ij}$ between $X_i$ and $X_j$. We can express the precision matrix $\bOmega$ as $\bOmega = \bmD^{1/2}\bGamma\bmD^{1/2},$ where $\bmD = \mathrm{diag}(\omega_{11},\ldots,\omega_{pp})$ and $\bGamma = (\!(\gamma_{ij})\!)$ is such that $\gamma_{ii} = 1, i = 1,\ldots,p,$ and $\gamma_{ij} = - \rho^{ij}, 1 \leq i \neq j \leq p.$ This gives,
\begin{equation}
    \label{eqn:precnparcor}
    \rho^{ij} = - \dfrac{\omega_{ij}}{\sqrt{\omega_{ii}\omega_{jj}}}, \, 1\leq i \neq j \leq p.
\end{equation}
From equation (\ref{eqn:precnparcor}), we get an alternate representation of the regression coefficient $\beta_{jk}$ as $\rho^{jk}\sqrt{\omega_{kk}/\omega_{jj}}$, so that the partial correlation $\rho^{ij}$ can be written as $\rho^{ij} = \mathrm{sign}(\beta_{ij})\sqrt{\beta_{ij}\beta_{ji}}.$ Also, note that the sparsity structure of $\bOmega$ and the corresponding partial correlation matrix $\bm{R} = (\!(\rho^{ij})\!)$ are identical. Thus, estimation of the elements of the precision matrix or the partial correlation matrix reduces to fitting $p$ parallel regression models to the individual excess asset returns with the remainder of the asset returns as predictors.

\subsection{Joint regression shrinkage}

We propose to work with the joint regression shrinkage approach named \texttt{space} \citep{peng2009} that considers a joint loss function given by
\begin{equation*}
    L_n(\thetavec, \omegavec, \bX) = \dfrac{1}{2}\left(\sum_{i=1}^{p} \eta_i\left\|\bX_i - \sum_{j \neq i}\beta_{ij}\bX_j \right\|^2\right),
\end{equation*}
where $\bX_i = (X_i^{(1)}, \ldots, X_i^{(n)})', \bX = \{\bX^{(t)}\}_{t=1}^{n}, \thetavec = (\rho^{12},\ldots,\rho^{(p-1)p})', \omegavec = (\omega_{11},\ldots, \omega_{pp})',$ and $\etavec = \{\eta_i\}_{i=1}^{p}$ are non-negative weights. The \texttt{space} method minimizes the joint penalized loss function

\begin{equation}
\label{eqn:jointloss}
\mathcal{L}_n(\thetavec, \omegavec, \bX) = L_n(\thetavec, \omegavec, \bX) + \lambda \sum_{1\leq i \neq j \leq p}|\rho^{ij}|,
\end{equation}
where $\lambda$ is the tuning parameter. The proposed approach for partial correlation estimation has several advantages over other nodewise regression-based approaches. First, since we are estimating the partial correlations directly using the penalized joint loss function as in (\ref{eqn:jointloss}), the sign consistency of the regression co-efficients $\beta_{ij}$ and $\beta_{ji}$ are preserved automatically, and the estimated precision matrix is also symmetric. These are in contrast to the nodewise regression methods of \cite{meinshausen2006} and \cite{callot2021nodewise}, where the sign consistency is not guaranteed, and the estimated precision matrix is not necessarily symmetric. Additionally, owing to direct estimation of the partial correlations and the residual variances, the number of unknown parameters to be estimated is $p(p+1)/2$ in comparison to $(p-1)^2$ for nodewise regression approaches, thus making the computations for the proposed method faster and more efficient. Finally, the weights $\etavec$ can be chosen accordingly to adjust for heteroscedastic noises in the regressions or account for the overall dependence structure among the asset returns.

\section{Simulations}
\label{sec:Simu}

We assess the performance of our proposed method under different simulated settings and also compare the results with three competing methods, namely, the nodewise regression method \citep{callot2021nodewise}, factor model based POET estimator \citep{fan2013large}, and the shrinkage based Ledoit-Wolf estimator \citep{ledoit2004well}. We consider both the unweighted and weighted versions (named \texttt{space-unweighted} and \texttt{space-weighted}) of the proposed procedure, as outlined in \cite{peng2009}. Following the simulation setup as described in \cite{callot2021nodewise}, we consider two data generation processes (DGPs), given by the Toeplitz structure for the covariance, and a sparse factor structure for returns. The details of the DGPs are provided in the Supplementary section. 

\subsection{Performance evaluation metrics}

We consider three different error metrics for evaluating the performance of the methods, namely, the weight estimation error, variance estimation error, and the risk estimation error, defined below.

We denote the optimal portfolio weight vector by $\wvec_O$, where $\wvec_O$ can be either $\wvec_G$ or $\wvec_M$ corresponding to the global minimum variance portfolio or the Markowitz portfolio as given in equations (\ref{eqn:GMVportfolio}) and (\ref{eqn:MPortfolio}) respectively. The estimated counterpart of $\wvec_O$, obtained via the plug-in estimate of $\bOmega$ is denoted by $\hat{\wvec}_O$. The portfolio weight estimation error $E_W$ is defined as the $L_1$-norm of the difference between $\hat{\wvec}_O$ and $\wvec_O$, that is, 
\[E_W = \|\hat{\wvec}_O - \wvec_O\|_1.\]
The variance estimation error $E_V$ is likewise defined as the absolute relative difference in the estimated optimal variance and the true optimal variance, that is, 
\[E_V = \left|\dfrac{\hat{\wvec}_O'\hat{\bSigma}\hat{\wvec}_O}{\wvec_O'\bSigma\wvec_O} - 1\right|.\]
Finally, the risk estimation error $E_R$ is defined as 
\[E_R = |\hat{\wvec}_O'(\hat{\bSigma} - \bSigma)\hat{\wvec}_O|.\]

\subsection{Simulation setting}

We consider varying sample sizes $n = 100, 200, 400$ with corresponding choices of the dimension $p$ to be $p = n/2 \ (n > p)$ and $p = 3n/2 \ (n < p)$ for all the different DGPs. For each of the $(n,p)$ combinations, we run 100 replications and report the performance metrics for the proposed joint shrinkage methods along with its competitors. We use the \texttt{space} package in \texttt{R} to implement the proposed method, and the \texttt{R} codes provided in the supplementary section of \cite{callot2021nodewise} for the competing methods. 

\subsection{Results}

The results under the different DGPs and varying combinations of $(n,p)$ are shown in Tables \ref{tbl1a}, \ref{tbl1b}, \ref{tbl2a}, and \ref{tbl2b} corresponding to the Toeplitz covariance structure and in Tables \ref{tbl3a}, \ref{tbl3b}, \ref{tbl4a}, and \ref{tbl4b} corresponding to the sparse factor structure for returns. The proposed joint shrinkage methods, both the unweighted and weighted versions, demonstrate excellent performance in all three metrics under the Toeplitz covariance structure. The variance and weight estimation errors for the joint shrinkage methods are better than both the nodewise and POET estimators, while the risk estimation errors are the least compared to all the competing methods. Under the sparse factor structure of the returns, the joint shrinkage methods and the Ledoit-Wolf method perform better than the other two methods. Overall, all the errors decrease with increasing sample size $n$, irrespective of the increasing dimension $p$. These results validate the utility of the proposed joint shrinkage approach for the portfolio optimization task, especially when the asset class is large.

\begin{table}[htbp]
\renewcommand{\thetable}{\arabic{table}a}
\caption{ Toeplitz DGP: global minimum variance portfolio, $p=n/2$} 
\label{tbl1a}
\begin{adjustbox}{width=\columnwidth,center}
\begin{tabular}{l c c c l c c c l c c c} 
\toprule 
\midrule
  & \multicolumn{3}{c}{$n=100,p=50$} 
& & \multicolumn{3}{c}{$n=200,p=100$}
& & \multicolumn{3}{c}{$n=400,p=200$}\\
\cmidrule(l){2-4} \cmidrule(l){6-8} \cmidrule(l){10-12}  
 & $E_V$ & $E_W$ & $E_R$ &  & $E_V$ & $E_W$ & $E_R$ &  & $E_V$ & $E_W$ & $E_R$   \\ 
\midrule 
space-unweighted&	
0.3860	& 0.1196	& 0.0034 &&		
0.3627	& 0.0852	& 0.0012	&&	
0.3406	& 0.0646	& 0.0004\\
space-weighted&	
0.3851	& 0.1180	& 0.0034	&&	
0.3674	& 0.0822	& 0.0012	&&	
0.3541	& 0.0596	& 0.0004\\
Nodewise &	
0.4013	& 0.2488 & 0.0038 &&		
0.3788	& 0.1718 & 0.0012	&&	
0.3624	& 0.1180 & 0.0003\\
POET &
0.5695	& 0.3874	& 0.0146	&&	
0.4222	& 0.2869	& 0.0054	&&	
0.3281	& 0.2061	& 0.0029\\
Ledoit–Wolf&	
0.3216	&0.0642	&0.0066	&&	
0.3173	&0.0602	&0.0033	&&	
0.3200	&0.0572	&0.0017\\
\midrule
\bottomrule 
\end{tabular}
\end{adjustbox}
\end{table}

\begin{table}[htbp]
\addtocounter{table}{-1}
\renewcommand{\thetable}{\arabic{table}b}
\caption{ Toeplitz DGP: global minimum variance portfolio, $p=3n/2$} 
\label{tbl1b}
\begin{adjustbox}{width=\columnwidth,center}
\begin{tabular}{l c c c l c c c l c c c} 
\toprule 
\midrule
  & \multicolumn{3}{c}{$n=100,p=150$} 
& & \multicolumn{3}{c}{$n=200,p=300$}
& & \multicolumn{3}{c}{$n=400,p=600$}\\
\cmidrule(l){2-4} \cmidrule(l){6-8} \cmidrule(l){10-12}  
 & $E_V$ & $E_W$ & $E_R$ &  & $E_V$ & $E_W$ & $E_R$ &  & $E_V$ & $E_W$ & $E_R$  \\ 
\midrule 
space-unweighted & 
0.3968 & 0.1155 & 0.0010	
&&	
0.3711 & 0.0824 & 0.0004	
&&	
0.3522 & 0.0614 & 0.0001\\
space-weighted &	
0.3935 & 0.1156 & 0.0010
&&
0.3724 & 0.0810 & 0.0004
&&		
0.3605 & 0.0580	& 0.0001\\
Nodewise & 
0.4185 & 0.2339 & 0.0013
&&
0.3883 & 0.1628 & 0.0004
&&
0.3697 & 0.1155 & 0.0001\\
POET & 
0.6616 & 0.3400 & 0.0048
&&
0.4617 & 0.2653 & 0.0018
&&
0.3237 & 0.2252 & 0.0007\\
Ledoit–Wolf & 
0.3492 & 0.0516 & 0.0023
&&
0.3429 & 0.0421 & 0.0011
&&
0.3421 & 0.0373 & 0.0006\\
\midrule
\bottomrule 
\end{tabular}
\end{adjustbox}
\end{table}

\begin{table}[htbp]
\renewcommand{\thetable}{\arabic{table}a}
\caption{ Toeplitz DGP: Markowitz portfolio,  $p=n/2$} 
\label{tbl2a}
\begin{adjustbox}{width=\columnwidth,center}
\begin{tabular}{l c c c l c c c l c c c} 
\toprule 
\midrule
  & \multicolumn{3}{c}{$n=100,p=50$} 
& & \multicolumn{3}{c}{$n=200,p=100$}
& & \multicolumn{3}{c}{$n=400,p=200$}\\
\cmidrule(l){2-4} \cmidrule(l){6-8} \cmidrule(l){10-12}  
 & $E_V$ & $E_W$ & $E_R$ &  & $E_V$ & $E_W$ & $E_R$ &  & $E_V$ & $E_W$ & $E_R$  \\ 
\midrule 
space-unweighted    & 0.3452 & 0.4145 & 0.0035 &  & 0.2727 & 0.2002 & 0.0012 &  & 0.2612 & 0.1616 & 0.0004 \\
space-weighted      & 0.3446 & 0.4150 & 0.0035 &  & 0.2751 & 0.1991 & 0.0012 &  & 0.2686 & 0.1597 & 0.0004 \\
Nodewise            & 0.3571 & 0.392  & 0.0118 &  & 0.2818 & 0.3031 & 0.0036 &  & 0.275  & 0.1875 & 0.0017 \\
POET                & 0.5074 & 0.4862 & 0.0148 &  & 0.3744 & 0.2688 & 0.0056 &  & 0.3298 & 0.2088 & 0.0024 \\
Ledoit–Wolf         & 0.3416 & 0.3921 & 0.0071 &  & 0.2722 & 0.1759 & 0.0035 &  & 0.2704 & 0.1494 & 0.0017 \\
\midrule
\bottomrule 
\end{tabular}
\end{adjustbox}
\end{table}

\begin{table}[htbp]
\addtocounter{table}{-1}
\renewcommand{\thetable}{\arabic{table}b}
\caption{ Toeplitz DGP: Markowitz portfolio, $p=3n/2$} 
\label{tbl2b}
\begin{adjustbox}{width=\columnwidth,center}
\begin{tabular}{l c c c l c c c l c c c} 
\toprule 
\midrule
  & \multicolumn{3}{c}{$n=100,p=150$} 
& & \multicolumn{3}{c}{$n=200,p=300$}
& & \multicolumn{3}{c}{$n=400,p=600$}\\
\cmidrule(l){2-4} \cmidrule(l){6-8} \cmidrule(l){10-12}  
 & $E_V$ & $E_W$ & $E_R$ &  & $E_V$ & $E_W$ & $E_R$ &  & $E_V$ & $E_W$ & $E_R$  \\ 
\midrule 
space-unweighted    & 0.2862 & 0.1716 & 0.0012 &  & 0.2723 & 0.1346 & 0.0004 &  & 0.2593 & 0.0751 & 0.0001   \\
space-weighted      & 0.2844 & 0.1714 & 0.0012 &  & 0.2729 & 0.1347 & 0.0004 &  & 0.2639 & 0.0722 & 0.0001   \\
Nodewise            & 0.2975 & 0.2836 & 0.0024 &  & 0.2825 & 0.1821 & 0.0013 &  & 0.2696 & 0.1086 & 0.0006 \\
POET                & 0.4405 & 0.3198 & 0.0048 &  & 0.3727 & 0.258  & 0.0019 &  & 0.3289 & 0.2001 & 0.0008 \\
Ledoit–Wolf         & 0.2693 & 0.1288 & 0.0024 &  & 0.2674 & 0.102  & 0.0018 &  & 0.2616 & 0.0425 & 0.0006 \\
\midrule
\bottomrule 
\end{tabular}
\end{adjustbox}
\end{table}

\begin{table}[htbp]
\renewcommand{\thetable}{\arabic{table}a}
\caption{Sparse factor DGP: global minimum variance portfolio, $p=n/2$} 
\label{tbl3a}
\begin{adjustbox}{width=\columnwidth,center}
\begin{tabular}{l c c c l c c c l c c c} 
\toprule 
\midrule
  & \multicolumn{3}{c}{$n=100,p=50$} 
& & \multicolumn{3}{c}{$n=200,p=100$}
& & \multicolumn{3}{c}{$n=400,p=200$}\\
\cmidrule(l){2-4} \cmidrule(l){6-8} \cmidrule(l){10-12}  
 & $E_V$ & $E_W$ & $E_R$ &  & $E_V$ & $E_W$ & $E_R$ &  & $E_V$ & $E_W$ & $E_R$  \\ 
\midrule 
space-unweighted    & 0.0368 & 0.1141 & 0.0025   &  & 0.0176 & 0.0812 & 0.0009   &  & 0.0083 & 0.0592 & 0.0003   \\
space-weighted      & 0.0337 & 0.1146 & 0.0025   &  & 0.0158 & 0.0811 & 0.0009   &  & 0.0077 & 0.0591 & 0.0003   \\
Nodewise            & 0.0760  & 0.2226 & 0.0027   &  & 0.0415 & 0.1605 & 0.0009 &  & 0.0179 & 0.1129 & 0.0003 \\
POET                & 0.3593 & 0.2886 & 0.008    &  & 0.1718 & 0.2626 & 0.0021   &  & 0.0895 & 0.1477 & 0.0006 \\
Ledoit–Wolf         & 0.0171 & 0.0188 & 0.0003 &  & 0.009  & 0.0155 & 0.0001 &  & 0.0042 & 0.0163 & 0.0001 \\
\midrule
\bottomrule 
\end{tabular}
\end{adjustbox}
\end{table}

\begin{table}[htbp]
\addtocounter{table}{-1}
\renewcommand{\thetable}{\arabic{table}b}
\caption{ Sparse factor DGP: global minimum variance portfolio, $p=3n/2$} 
\label{tbl3b}
\begin{adjustbox}{width=\columnwidth,center}
\begin{tabular}{l c c c l c c c l c c c} 
\toprule 
\midrule
  & \multicolumn{3}{c}{$n=100,p=150$} 
& & \multicolumn{3}{c}{$n=200,p=300$}
& & \multicolumn{3}{c}{$n=400,p=600$}\\
\cmidrule(l){2-4} \cmidrule(l){6-8} \cmidrule(l){10-12}  
 & $E_V$ & $E_W$ & $E_R$ &  & $E_V$ & $E_W$ & $E_R$ &  & $E_V$ & $E_W$ & $E_R$  \\ 
\midrule 
space-unweighted    & 0.0359 & 0.1140 & 0.0007   &  & 0.0166 & 0.0820 & 0.0003   &  & 0.0077 & 0.0607 & 0.00008 \\
space-weighted      & 0.0320 & 0.1150 & 0.0007   &  & 0.0149 & 0.0822 & 0.0003   &  & 0.0070 & 0.0607 & 0.00008 \\
Nodewise            & 0.0800 & 0.2285 & 0.001    &  & 0.0373 & 0.1617 & 0.0003 &  & 0.0159 & 0.1148 & 0.00009 \\
POET                & 0.3221 & 0.2921 & 0.0025   &  & 0.1762 & 0.2331 & 0.0008 &  & 0.1008 & 0.196  & 0.00030 \\
Ledoit–Wolf         & 0.0138 & 0.0225 & 0.0001 &  & 0.0065 & 0.0209 & 0.00002 &  & 0.0027 & 0.0198 & 0.00001 \\
\midrule
\bottomrule 
\end{tabular}
\end{adjustbox}
\end{table}

\begin{table}[htbp]
\renewcommand{\thetable}{\arabic{table}a}
\caption{Sparse factor DGP: Markowitz portfolio, $p=n/2$} 
\label{tbl4a}
\begin{adjustbox}{width=\columnwidth,center}
\begin{tabular}{l c c c l c c c l c c c} 
\toprule 
\midrule
  & \multicolumn{3}{c}{$n=100,p=50$} 
& & \multicolumn{3}{c}{$n=200,p=100$}
& & \multicolumn{3}{c}{$n=400,p=200$}\\
\cmidrule(l){2-4} \cmidrule(l){6-8} \cmidrule(l){10-12}  
 & $E_V$ & $E_W$ & $E_R$ &  & $E_V$ & $E_W$ & $E_R$ &  & $E_V$ & $E_W$ & $E_R$  \\ 
\midrule 
space-unweighted    & 0.1083 & 0.2879 & 0.0026 &  & 0.0245 & 0.1194 & 0.0009 &  & 0.0120 & 0.0806 & 0.00031 \\
space-weighted      & 0.1053 & 0.2882 & 0.0026 &  & 0.0231 & 0.1193 & 0.0009 &  & 0.0114 & 0.0805 & 0.00030 \\
Nodewise            & 0.1714 & 0.4596 & 0.0027 &  & 0.0367 & 0.2425 & 0.0009 &  & 0.0279 & 0.1858 & 0.00030 \\
POET                & 0.3066 & 0.3928 & 0.0085 &  & 0.1509 & 0.2210  & 0.0021 &  & 0.0821 & 0.1573 & 0.00060 \\
Ledoit–Wolf         & 0.0866 & 0.2592 & 0.0004 &  & 0.0166 & 0.0890  & 0.0001 &  & 0.0081 & 0.0551 & 0.00002 \\
\midrule
\bottomrule 
\end{tabular}
\end{adjustbox}
\end{table}

\begin{table}[htbp]
\addtocounter{table}{-1}
\renewcommand{\thetable}{\arabic{table}b}
\caption{ Sparse factor DGP: Markowitz portfolio, $p=3n/2$} 
\label{tbl4b}
\begin{adjustbox}{width=\columnwidth,center}
\begin{tabular}{l c c c l c c c l c c c} 
\toprule 
\midrule
  & \multicolumn{3}{c}{$n=100,p=150$} 
& & \multicolumn{3}{c}{$n=200,p=300$}
& & \multicolumn{3}{c}{$n=400,p=600$}\\
\cmidrule(l){2-4} \cmidrule(l){6-8} \cmidrule(l){10-12}  
 & $E_V$ & $E_W$ & $E_R$ &  & $E_V$ & $E_W$ & $E_R$ &  & $E_V$ & $E_W$ & $E_R$  \\ 
\midrule 
space-unweighted    & 0.0364 & 0.1329 & 0.0008 &  & 0.0166 & 0.0942 & 0.00030 &  & 0.0074 & 0.0620 & 0.00008 \\
space-weighted      & 0.0329 & 0.1340 & 0.0008 &  & 0.0151 & 0.0945 & 0.00030 &  & 0.0067 & 0.0620 & 0.00008 \\
Nodewise            & 0.0586 & 0.2618 & 0.0009 &  & 0.0294 & 0.1901 & 0.00030 &  & 0.0098 & 0.1234 & 0.00009 \\
POET                & 0.2396 & 0.3029 & 0.0025 &  & 0.1473 & 0.2354 & 0.00080 &  & 0.0925 & 0.1961 & 0.00030 \\
Ledoit–Wolf         & 0.0157 & 0.0739 & 0.0001 &  & 0.0075 & 0.0484 & 0.00002 &  & 0.0028 & 0.0246 & 0.00001 \\
\midrule
\bottomrule 
\end{tabular}
\end{adjustbox}
\end{table}


\section{Real datasets}
\label{sec:Real}

We now demonstrate the performance of our proposed method and the competing approaches on real datasets of excess asset returns. We first describe the metrics on the basis of which we will compare the performances of the different methods.

\subsection{Performance evaluation metrics}

We adopt a `rolling-sample' approach \citep{demiguel2009optimal} to evaluate the performances of the portfolio optimization methods. For $p$-dimensional excess returns data $\bX^{(1)},\ldots,\bX^{(n)}$ spanning $n$ time points (daily or monthly), we split the same in two portions, namely, the training and the testing data with respective lengths $n_t$ and $n - n_t.$ The portfolios are rebalanced at the start of each rolling-sample. The details of the rolling-sample approach are outlined in the Supplementary section. 

We compute four common metrics for performance evaluation -- out-of-sample mean return and variance, out-of-sample Sharpe ratio, and the portfolio turnover. We also consider portfolios including and excluding associated transaction costs. The out-of-sample Sharpe Ratio is given by $SR = \hat{\mu}/\hat{\sigma},$ where $\hat{\mu}$ and $\hat{\sigma}^2$ are respectively the out-of-sample mean portfolio return and variance. In the absence of transaction costs, these quantities are defined as:
\[
\begin{aligned}
\hat{\mu} = & \frac{1}{n - n_t}\sum_{t = n_t}^{n-1}\hat{\wvec}_{t}'\bX^{(t+1)},\;
\hat{\sigma}^2 = & \frac{1}{(n - n_t) - 1}\sum_{t=n_t}^{n-1}\left(\hat{\wvec}_{t}'\bX^{(t+1)} - \hat{\mu}\right)^2,
\end{aligned}
\]
where $\hat{\wvec}_t$ is the estimated portfolio weight at time $t$. In the presence of transaction cost $c$, we need to modify the above-defined measures suitably. Denoting $\hat{\wvec}_{j,t}^+$ as the estimated portfolio weight before rebalancing at time point $t+1$, the net return at $t+1$ accounting for the transaction cost is given by 
\[
R_{t+1} = \hat{\wvec}_{t}'\bX^{(t+1)} - c(1+\hat{\wvec}_{t}'\bX^{(t+1)}) \sum_{j=1}^{p}\left|\hat{\wvec}_{t+1,j} - \hat{\wvec}^{+}_{t,j}\right|.
\]
The modified out-of-sample mean and variance of the portfolio under transaction cost $c$ are then given by,
\[
\begin{aligned}
\hat{\mu}_{c} = & \frac{1}{n - n_t}\sum_{t = n_t}^{n-1}R_{t+1},\;
\hat{\sigma}^2_{c} = & \frac{1}{(n - n_t) - 1}\sum_{t=n_t}^{n-1}\left(R_{t+1} - \hat{\mu}_{c}\right)^2.
\end{aligned}
\]
The Sharpe Ratio is modified suitably as $SR_c = \hat{\mu}_c/\hat{\sigma}_c$. Under presence of transaction cost, the portfolio turnover is defined as $\mathrm{Turnover} = (n - n_t)^{-1}\sum_{t = n_t}^{n-1} \sum_{j=1}^{p} \left|\hat{\wvec}_{t+1,j} - \hat{\wvec}^{+}_{t,j}\right|.$ 

\subsection{Datasets}

We have used daily and monthly returns of components of the S\&P 500 index. With the monthly data, we have $n < p$ setup, while for daily data we have $p < n$. The data are split into training and testing sets, and thereafter we use the rolling-sample approach as described above. A brief description of the data, including the training and the testing periods, are provided in Table \ref{tbl:datadesrp}. We use monthly and daily equivalent targets corresponding to $10\%$ compounded annual return for the Markowitz portfolio, and take the three-month treasury bill rate as the risk-free return rate. The proportional transaction cost is set to $c = 50$ basis points per transaction, as in \cite{demiguel2009optimal}.

\begin{table}[ht]
\caption{Data Description} 
\label{tbl:datadesrp}
\begin{adjustbox}{width=\columnwidth,center}
\begin{tabular}{cccccc}
\toprule 
\midrule
\multicolumn{6}{c}{\textit{Monthly data: January 1994 -- May 2018, $p$ = 304}} \\
\multicolumn{6}{c}{\textit{Markowitz target return: $0.7974\%$}} \\
\multicolumn{2}{c}{Training data} & \multicolumn{2}{c}{Test data} &                   &                        \\
Start date         & End date          & Start date       & End date        & \#traning samples & \#test samples \\
\midrule
Jan, 1994     & Mar, 2010    & Apr, 2010   & May, 2018  & 195               & 98             \\
Jan, 1994     & May, 2008    & Jun, 2008   & May, 2018  & 173               & 120             \\
\midrule
\multicolumn{6}{c}{\textit{Daily data: 02 July 1994 -- 30 April 2018, $p$ = 452}} \\
\multicolumn{6}{c}{\textit{Markowitz target return: $0.0378\%$}} \\
\multicolumn{2}{c}{Training data} & \multicolumn{2}{c}{Test data} &                   &                        \\
Start date         & End date          & Start date       & End date        & \#traning samples & \#test samples \\
\midrule
02 Jul, 2013     & 28 Apr, 2017    & 01 May, 2017   & 30 Apr, 2018  & 964               & 252             \\
02 Jul, 2013     & 29 Jan, 2018    & 30 Jan, 2018   & 30 Apr, 2018  & 1153               & 63             \\
\midrule
\bottomrule
\end{tabular}
\end{adjustbox}
\end{table}

\begin{table}[ht]
\caption{Portfolio Optimization performance for monthly and daily returns data.} 
\label{rtbl1}
\begin{adjustbox}{width=\columnwidth,center}
\begin{tabular}{r c c c c c c ccccrcccrcc} 
\toprule 
\midrule
 & &\multicolumn{7}{c}{Global minimum portfolio}& & &
\multicolumn{7}{c}{Markowitz portfolio}\\
\cmidrule(l){2-9} \cmidrule(l){12-18}
 & & Return & & Variance & & SR & & Turnover & & & 
Return & & Variance & & SR & &Turnover\\ 
\midrule 
\multicolumn{18}{c}{\textit{Monthly returns. Training: Jan, 1994 -- Mar, 2010, Testing: Apr, 2010 -- May, 2018}} \\
\textbf{\textit{Without transaction cost}} &&&&&&&&&& \\
space–unweighted    &  & 0.02907 &  & 0.01931 &  & 0.2092 &  & – &  &  & 0.02438  &  & 0.03828 &  & 0.1246  &  & – \\
space–weighted      &  & 0.02888 &  & 0.02459 &  & 0.1841 &  & – &  &  & 0.03437  &  & 0.04342 &  & 0.1650  &  & – \\
Nodewise            &  & 0.02644 &  & 0.01580  &  & 0.2104 &  & – &  &  & 0.03503  &  & 0.04585 &  & 0.1635  &  & – \\
POET                &  & 0.02499 &  & 0.01953 &  & 0.1788 &  & – &  &  & 0.03317  &  & 0.05248 &  & 0.1448  &  & – \\
Ledoit–Wolf         &  & 0.07140 &  & 0.18421 &  & 0.1664 &  & – &  &  & –0.01128 &  & 0.24568 &  & –0.0227 &  & – \\
\midrule
\textbf{\textit{With transaction cost}} &&&&&&&&&& \\
space–unweighted    &  & 0.02859 &  & 0.01946 &  & 0.2050 &  & 0.0700 &  &  & 0.02343  &  & 0.03859 &  & 0.1193  &  & 0.1171 \\
space–weighted      &  & 0.02903 &  & 0.02474 &  & 0.1846 &  & 0.0642 &  &  & 0.03458  &  & 0.04369 &  & 0.1654  &  & 0.1088 \\
Nodewise            &  & 0.02650 &  & 0.01590  &  & 0.2102 &  & 0.1268 &  &  & 0.03517  &  & 0.04613 &  & 0.1637  &  & 0.1709 \\
POET                &  & 0.02497 &  & 0.01964 &  & 0.1782 &  & 0.2080 &  &  & 0.03324  &  & 0.05274 &  & 0.1447  &  & 0.2497 \\
Ledoit–Wolf         &  & 0.07175 &  & 0.18545 &  & 0.1666 &  & 0.2150 &  &  & –0.01166 &  & 0.24733 &  & –0.0234 &  & 0.4421 \\
\midrule
\multicolumn{18}{c}{\textit{Daily returns. Training: 02 Jul, 2013 -- 28 Apr, 2017, Testing: 01 May, 2017 -- 30 Apr, 2018}} \\
\textbf{\textit{Without transaction cost}} &&&&&&&&&& \\
space–unweighted    &  & 4.094$e$–04 &  & 4.47$e$–05 &  & 0.0613 &  & – &  &  & 3.435$e$–04  &  & 4.37$e$–05 &  & 0.0519 &  & – \\
space–weighted      &  & 3.560$e$–04 &  & 3.69$e$–05  &  & 0.0586 &  & – &  &  & 3.049$e$–04  &  & 3.60$e$–05 &  & 0.0508 &  & – \\
Nodewise            &  & 4.424$e$–04 &  & 4.50$e$–05 &  & 0.0659 &  & – &  &  & 3.731$e$–04 &  & 4.42$e$–05 &  & 0.0561 &  & – \\
POET                &  & 4.537$e$–04 &  & 4.76$e$–05 &  & 0.0657 &  & – &  &  & 3.757$e$–04 &  & 4.63$e$–05 &  & 0.0552 &  & – \\
Ledoit–Wolf         &  & 4.103$e$–04 &  & 3.57$e$–05 &  & 0.0687 &  & – &  &  & 5.329$e$–04 &  & 3.74$e$–05 &  & 0.0871 &  & – \\
\midrule
\textbf{\textit{With transaction cost}} &&&&&&&&&& \\
space–unweighted    &  & 4.374$e$–04 &  & 4.45$e$–05 &  & 0.0656 &  & 0.0081 &  &  & 5.042$e$–04 &  & 4.36$e$–05 &  & 0.0530 &  & 0.0306 \\
space–weighted      &  & 3.404$e$–04 &  & 3.69$e$–05 &  & 0.0560 &  & 0.0414 &  &  & 2.748$e$–04 &  & 3.60$e$–05 &  & 0.0458 &  & 0.0570 \\
Nodewise            &  & 4.161$e$–04 &  & 4.49$e$–05 &  & 0.0620 &  & 0.0570 &  &  & 3.319$e$–04 &  & 4.42$e$–05 &  & 0.0499 &  & 0.0721 \\
POET                &  & 4.225$e$–04 &  & 4.77$e$–05 &  & 0.0612 &  & 0.0685 &  &  & 3.268$e$–04 &  & 4.63$e$–05 &  & 0.0480 &  & 0.0884 \\
Ledoit–Wolf         &  & 4.174$e$–05 &  & 3.56$e$–05 &  & 0.0069 &  & 0.3430 &  &  & 1.490$e$–04 &  & 3.77$e$–05 &  & 0.0242 &  & 0.3539 \\
\midrule 
\bottomrule 
\end{tabular}
\end{adjustbox}
\end{table}
\subsection{Results}

The performances of the different methods are presented in Table~\ref{rtbl1} corresponding to the first training periods. For the monthly returns data, the Ledoit-Wolf estimator yields the highest returns corresponding to the global minimum variance portfolios, but at the cost of substantial risks, resulting in poor Sharpe ratios. The performances of the proposed joint shrinkage-based methods are generally the best, or at least comparable with the nodewise regression approach. For the Markowitz portfolio, our proposed methods show the best performance in terms of the best returns, lowest risks and highest Sharpe ratios. Additionally, in the presence of transaction costs, the joint shrinkage methods give the lowest turnover, thus preventing higher fees to adjust for the higher turnover rates. For the daily returns data, they perform the best with respect to all the metrics when transaction cost is present. These results validate the utility of joint shrinkage in the portfolio optimization problem. We present additional real data results corresponding to the second set of training samples in the Supplementary section.

\section{Discussion}

In this paper, we have proposed using a regression-based joint shrinkage method to estimate the partial correlations and residual variances for a large number of assets. The estimated parameters are then used to solve a portfolio optimization problem, corresponding to two optimal portfolio choices. Through extensive numerical studies, we demonstrate the excellent performances of our proposed approach as compared to competing methods. Our proposed approach also enjoys the ease of implementation and the availability of fast computational methods, even for a large number of financial assets.

\section*{Acknowledgements}
S.B. is supported by IIM Indore Young Faculty Research Chair Award grant.

\bibliographystyle{apalike}
\bibliography{cite.bib}

\begin{thebibliography}{}

\bibitem[Callot et~al., 2021]{callot2021nodewise}
Callot, L., Caner, M., {\"O}nder, A.~{\"O}., and Ula{\c{s}}an, E. (2021).
\newblock A nodewise regression approach to estimating large portfolios.
\newblock {\em Journal of Business \& Economic Statistics}, 39(2):520--531.

\bibitem[DeMiguel et~al., 2009]{demiguel2009optimal}
DeMiguel, V., Garlappi, L., and Uppal, R. (2009).
\newblock Optimal versus naive diversification: How inefficient is the 1/n
  portfolio strategy?
\newblock {\em The Review of Financial Studies}, 22(5):1915--1953.

\bibitem[Fan et~al., 2013]{fan2013large}
Fan, J., Liao, Y., and Mincheva, M. (2013).
\newblock Large covariance estimation by thresholding principal orthogonal
  complements.
\newblock {\em Journal of the Royal Statistical Society. Series B, Statistical
  methodology}, 75(4).

\bibitem[Friedman et~al., 2008]{friedman2008sparse}
Friedman, J., Hastie, T., and Tibshirani, R. (2008).
\newblock Sparse inverse covariance estimation with the graphical lasso.
\newblock {\em Biostatistics}, 9(3):432--441.

\bibitem[Ledoit and Wolf, 2004]{ledoit2004well}
Ledoit, O. and Wolf, M. (2004).
\newblock A well-conditioned estimator for large-dimensional covariance
  matrices.
\newblock {\em Journal of Multivariate Analysis}, 88(2):365--411.

\bibitem[Markowitz, 1952]{markowitz1952}
Markowitz, H. (1952).
\newblock Portfolio selection.
\newblock {\em The Journal of Finance}, 7(1):77--91.

\bibitem[Meinshausen and Bühlmann, 2006]{meinshausen2006}
Meinshausen, N. and Bühlmann, P. (2006).
\newblock {High-dimensional graphs and variable selection with the Lasso}.
\newblock {\em The Annals of Statistics}, 34(3):1436 -- 1462.

\bibitem[Peng et~al., 2009]{peng2009}
Peng, J., Wang, P., Zhou, N., and Zhu, J. (2009).
\newblock Partial correlation estimation by joint sparse regression models.
\newblock {\em Journal of the American Statistical Association},
  104(486):735--746.
\newblock PMID: 19881892.

\bibitem[Sabharwal and Anjum, 2018]{sabharwal2018robo}
Sabharwal, C.~L. and Anjum, B. (2018).
\newblock Robo-revolution in the financial sector.
\newblock In {\em 2018 International Conference on Computational Science and
  Computational Intelligence (CSCI)}, pages 1289--1292. IEEE.

\end{thebibliography}

\newpage


\newcommand{\beginsupplement}{%

\setcounter{equation}{0}
\setcounter{page}{1}
\setcounter{table}{0}
\setcounter{section}{0}
\setcounter{subsection}{0}
\setcounter{figure}{0}
\renewcommand{\theequation}{S.\arabic{equation}}
\renewcommand{\thesection}{S.\arabic{section}}
\renewcommand{\thesubsection}{S.\arabic{subsection}}
\renewcommand{\thepage}{S.\arabic{page}}
\renewcommand{\thetable}{S.\arabic{table}}
\renewcommand{\thefigure}{S.\arabic{figure}}
}

\section*{Supplementary Material}
\beginsupplement
\label{sec:Supp}

\section{Data generating process details}
We have considered two data generating processes (DGPs) in our simulation studies, following the setup in \cite{callot2021nodewise}. We describe the DGPs in detail, along with the choice of underlying parameters.

\begin{enumerate}
    \item[(a)] \textit{Toeplitz structure for covariance}: For each time-point $t \in \{1,\cdots,n\}$, the vector of excess asset returns $\bX^{(t)} = \left(X_{1}^{(t)},\ldots,X_{p}^{(t)}\right)'$ is simulated from a $p$-dimensional Normal distribution with mean vector $\muvec$ and covariance matrix $\bSigma = (\!(\sigma_{ij})\!)$ where $\sigma_{ij} = 0.15^{|i-j|}$ for all $i,j \in \{1,\cdots,p\}$. We consider two different choices for the mean vector $\muvec$: (i) $\muvec = \zerovec$, and (ii) $\muvec \sim N_p(\zerovec, 10^{-4}\bm{I}_p).$ We evaluate the global minimum variance portfolio for the first choice of $\muvec$ and the Markowitz portfolio for the latter. Considering a $10\%$ annual return over 252 trading days, the target return is set at $\mu^* = 0.000376$.

\item[(b)] \textit{Sparse factor structure for returns}: We consider a weakly sparse three-factor structure for the excess returns, given by
\begin{equation*}
    \label{eqn:sparseDGP}
        \bX^{(t)} = \muvec + \bmB_t \bm{f}_t + \vepsvec_t,
    \end{equation*}
where $\bm{f}_t$ is a three-dimensional factor vector, $\bmB_t$ is a $p \times 3$ matrix of factor loadings, and $\vepsvec_t \sim N_p(\zerovec, \bmI_p)$ is the error term. The columns of $\bmB_t$ are generated independently from a $N_3(\zerovec, 10^{-2}\bmI_3)$ distribution, whereas the factors $\bm{f}_t$ are generated from a $N_3(\zerovec, 10^{-1}\bmI_3)$ distribution. Likewise the previous DGP, we consider two different choices for the mean vector $\muvec$ as: (i) $\muvec = \zerovec$, and (ii) $\muvec \sim N_p(\zerovec, 10^{-2}\bmI_p)$. We evaluate the performances of the different methods under the global minimum variance portfolio for the first choice of $\muvec$ and the Markowitz portfolio for the latter, with the same choice of the target return as in the previous scenario. 
\end{enumerate}

\section{Rolling-sample approach for evaluating performance in real datasets}

As mentioned in Section~5, we adopt the `rolling-sample' approach for evaluating the performances of the different portfolio optimization methods. We briefly describe the steps below:

\begin{enumerate}
    \item[(i)] First, the optimal portfolio weights $\hat{\wvec}_{n_t}$ are computed using the training data for the period $1,\ldots,n_t$, with the corresponding estimated portfolio return at the point $n_t + 1$ $\hat{\wvec}_{n_t}'\bX^{(n_t+1)}.$
    
    \item[(ii)] Next, the training window is rolled by one time point taking the data for the period $2,\ldots,n_t+1$ as the new training data to obtain the optimal weights along with the corresponding estimated return at time point $n_t + 2.$
    \item[(iii)] We iterate the above step till the end of the sample, computing the optimal weights taking $n-n_t,\ldots,n-1$ as the training data and estimating the return at time point $n.$ After each step above, we rebalance the portfolios at the start of the next time point.
\end{enumerate}

\section{Additional real data results}

We present the results for the different methods corresponding to the second set of training samples as described in Table~\ref{tbl:datadesrp}. The results are summarized in Table~\ref{rtbl2}.


\begin{table}[ht]
\caption{Portfolio Optimization performance for monthly and daily returns data.} 
\label{rtbl2}
\begin{adjustbox}{width=\columnwidth,center}
\begin{tabular}{r c c c c c c ccccrcccrcc} 
\toprule 
\midrule
 & &\multicolumn{7}{c}{Global minimum portfolio}& & &
\multicolumn{7}{c}{Markowitz portfolio}\\
\cmidrule(l){2-9} \cmidrule(l){12-18}
 & & Return & & Variance & & SR & & Turnover & & & 
Return & & Variance & & SR & &Turnover\\ 
\midrule 
\multicolumn{18}{c}{\textit{Monthly returns. Training: Jan, 1994 -- May, 2008, Testing: Jun, 2008 -- May, 2018}} \\
\textbf{\textit{Without transaction cost}} &&&&&&&&&& \\
space–unweighted    &  & 0.02300 &  & 0.01490 &  & 0.1884 &  & – &  &  & 0.022317  &  & 0.01949 &  & 0.1599  &  & – \\
space–weighted      &  & 0.02209 &  & 0.01647 &  & 0.1721 &  & – &  &  & 0.023685  &  & 0.02036 &  & 0.1660  &  & – \\
Nodewise            &  & 0.02158 &  & 0.01254 &  & 0.1927 &  & – &  &  & 0.024571  &  & 0.02054 &  & 0.1714  &  & – \\
POET                &  & 0.02078 &  & 0.01805 &  & 0.1547 &  & – &  &  & 0.022689  &  & 0.02537 &  & 0.1424  &  & – \\
Ledoit–Wolf         &  & 0.05039 &  & 0.11844 &  & 0.1464 &  & – &  &  & –0.009022 &  & 0.13209 &  & –0.0248 &  & – \\
\midrule
\textbf{\textit{With transaction cost}} &&&&&&&&&& \\
space–unweighted    &  & 0.02279 &  & 0.01496 &  & 0.1863 &  & 0.0554 &  &  & 0.021495  &  & 0.01958 &  & 0.1536 &  & 0.1150 \\
space–weighted      &  & 0.02215 &  & 0.01654 &  & 0.1722 &  & 0.0653 &  &  & 0.023744  &  & 0.02045 &  & 0.1660 &  & 0.1093 \\
Nodewise            &  & 0.02156 &  & 0.01261 &  & 0.192  &  & 0.1357 &  &  & 0.024580  &  & 0.02064 &  & 0.1710 &  & 0.1759 \\
POET                &  & 0.02070 &  & 0.01813 &  & 0.1538 &  & 0.1806 &  &  & 0.022634  &  & 0.02548 &  & 0.1418 &  & 0.2232 \\
Ledoit–Wolf         &  & 0.05041 &  & 0.11904 &  & 0.1461 &  & 0.2795 &  &  & –0.009442 &  & 0.13282 &  & –0.0259 &  & 1.8899 \\
\midrule
\multicolumn{18}{c}{\textit{Daily returns. Training: 02 Jul, 2013 -- 29 Jan, 2018, Testing: 30 Jan, 2018 -- 30 Apr, 2018}} \\
\textbf{\textit{Without transaction cost}} &&&&&&&&&& \\
space–unweighted    &  &–8.879$e$–04 &  & 1.294$e$–04 &  &–0.0780 &  & – &  &  &–0.000939 &  & 1.175$e$–04 &  &–0.0866 &  & – \\
space–weighted      &  &–8.487$e$–04 &  & 1.046$e$–04 &  &–0.0830 &  & – &  &  &–0.000933 &  & 1.004$e$–04 &  &–0.0932 &  & – \\
Nodewise            &  & –8.794$e$–04 &  & 1.262$e$–04 &  & –0.0782 &  & – &  &  & –0.000934 &  & 1.187$e$–04 &  & –0.0857 &  & – \\
POET                &  & –9.237$e$–04 &  & 1.330$e$–04 &  & –0.0800 &  & – &  &  & –0.001035 &  & 1.217$e$–04 &  & –0.0938 &  & – \\
Ledoit–Wolf         &  & –8.008$e$–04 &  & 5.874$e$–05 &  & –0.1044 &  & – &  &  & –0.000718 &  & 6.119$e$–05 &  & –0.0918 &  & – \\
\midrule
\textbf{\textit{With transaction cost}} &&&&&&&&&& \\
space–unweighted    &  &–7.721$e$–04 &  & 1.305$e$–04 &  &–0.0676 &  & 0.0092 &  &  &–0.000840 &  & 1.19$e$–04 &  &–0.0772 &  & 0.0351 \\
space–weighted      &  &–8.088$e$–04 &  & 1.058$e$–04 &  &–0.0786 &  & 0.0446 &  &  &–0.000904 &  & 1.02$e$–04 &  &–0.0897 &  & 0.0603 \\
Nodewise            &  & –8.323$e$–04 &  & 1.276$e$–04 &  & –0.0736 &  & 0.0540  &  &  & –0.000901 &  & 1.20$e$–04  &  & –0.0823 &  & 0.0709 \\
POET                &  & –8.059$e$–04 &  & 1.341$e$–04 &  & –0.0695 &  & 0.0110  &  &  & –0.000932 &  & 1.23$e$–04  &  & –0.0842 &  & 0.0371 \\
Ledoit–Wolf         &  & –1.258$e$–03 &  & 5.942$e$–05 &  & –0.1632 &  & 0.3269 &  &  & –0.001210 &  & 6.22$e$–05  &  & –0.1534 &  & 0.3419 \\
\midrule
\bottomrule 
\end{tabular}
\end{adjustbox}
\end{table}

For the second set of monthly returns training samples, we have similar conclusions regarding the performances of the different methods as in the first set. For the daily returns data, all the returns come out to be negative in the second training sample. The Ledoit-Wolf method gives the lowest return but with the lowest risks, but the joint shrinkage methods in almost all of the different scenarios result in the highest Sharpe ratios and lowest turnovers. 

\section{Visualizations}

\begin{figure}
\begin{subfigure}{\linewidth}
\centering
\includegraphics[width = 0.5\textwidth]{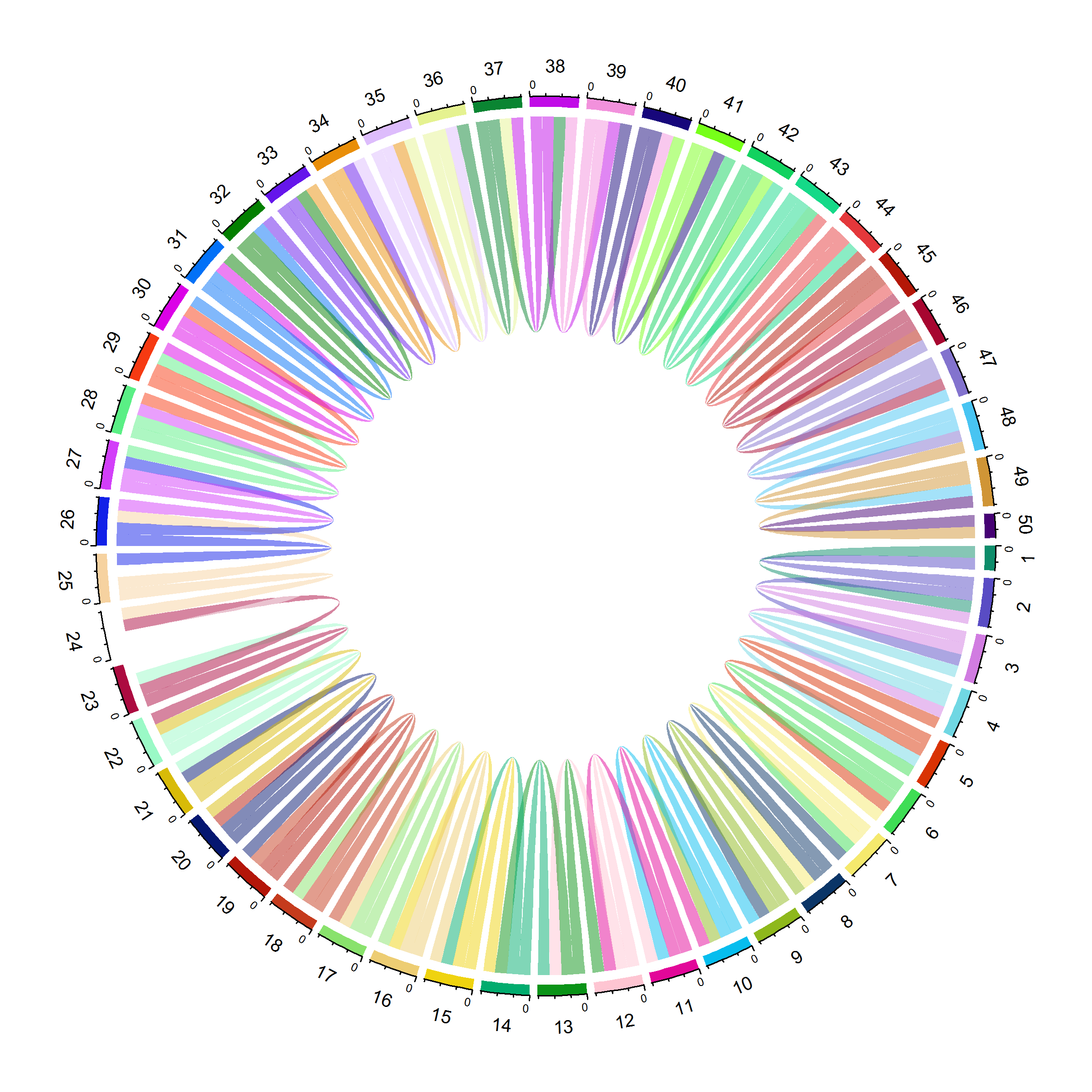}
\caption{true}
\label{fig1.1}
\end{subfigure}\\
\begin{subfigure}{0.5\linewidth}
\centering
\includegraphics[width = \textwidth]{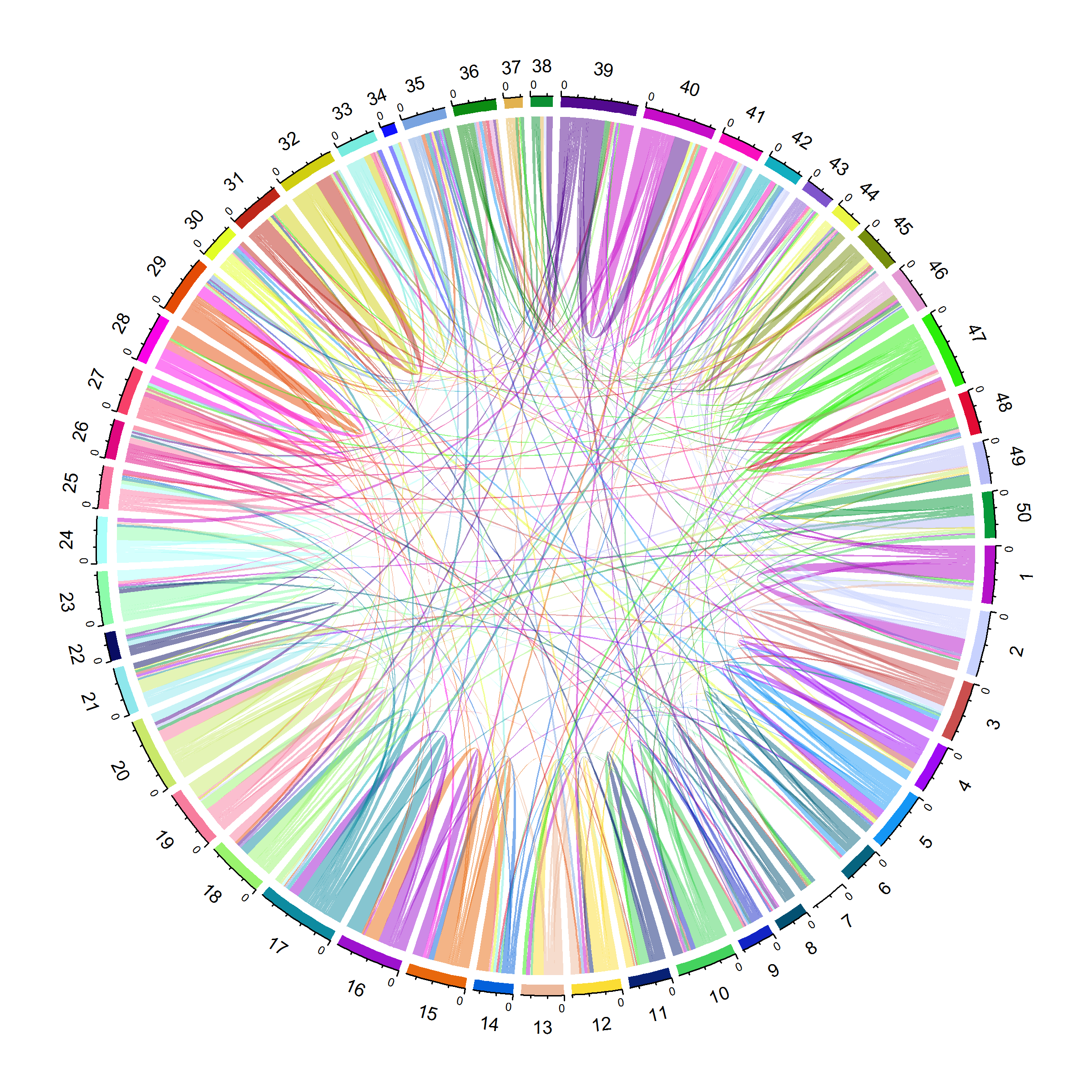}
\caption{nodewise}
\label{fig1.2}
\end{subfigure}
\begin{subfigure}{0.5\linewidth}
\centering
\includegraphics[width = \textwidth]{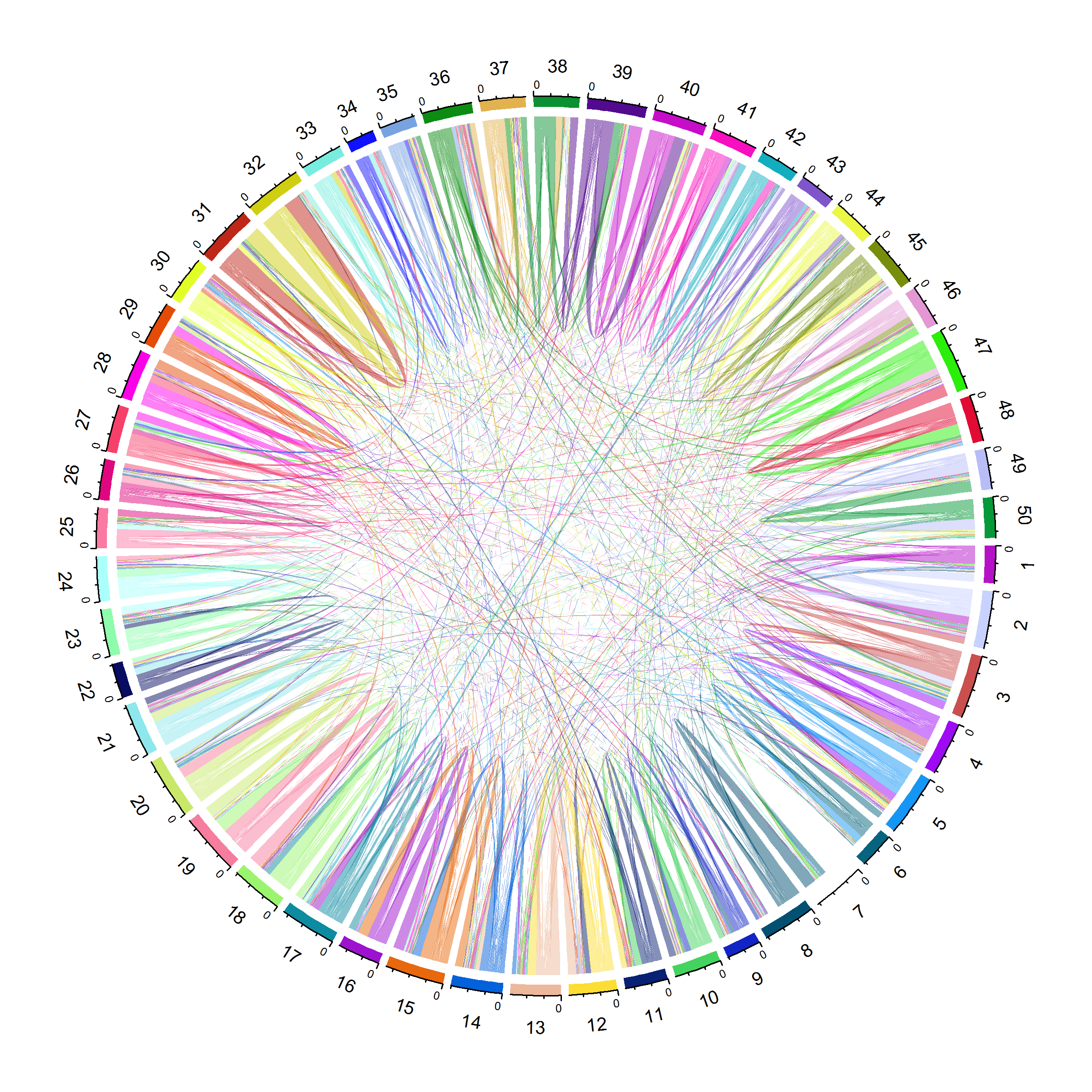}
\caption{space unweighted}
\label{fig1.3}
\end{subfigure}
\caption{Chord diagrams of true and estimated precision matrices corresponding to Toeplitz structure, $n = 100, p = 50$.}
\label{fig:fig1}
\end{figure}

\begin{figure}
\begin{subfigure}{\linewidth}
\centering
\includegraphics[width = 0.5\textwidth]{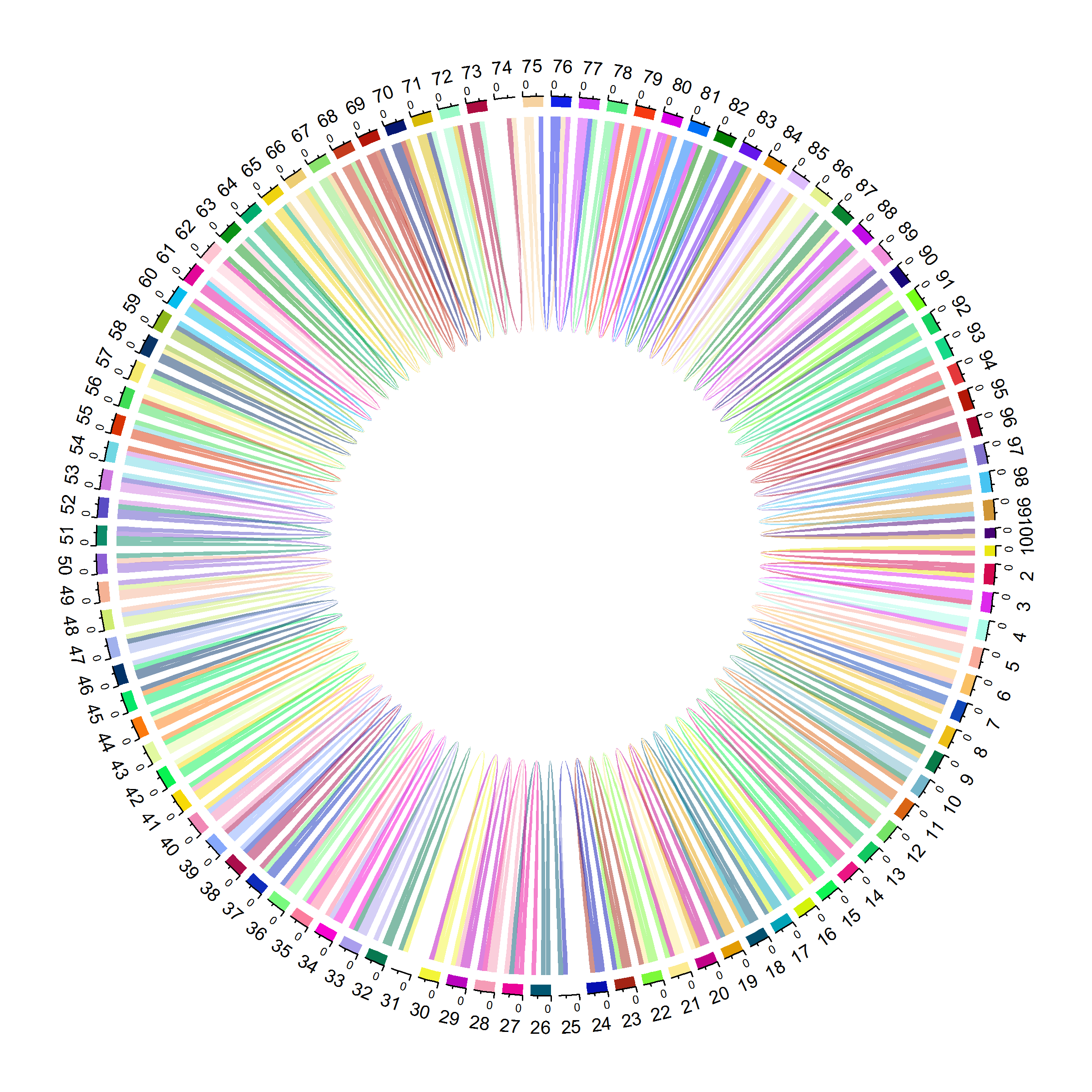}
\caption{true}
\label{fig2.1}
\end{subfigure}\\
\begin{subfigure}{0.5\linewidth}
\centering
\includegraphics[width = \textwidth]{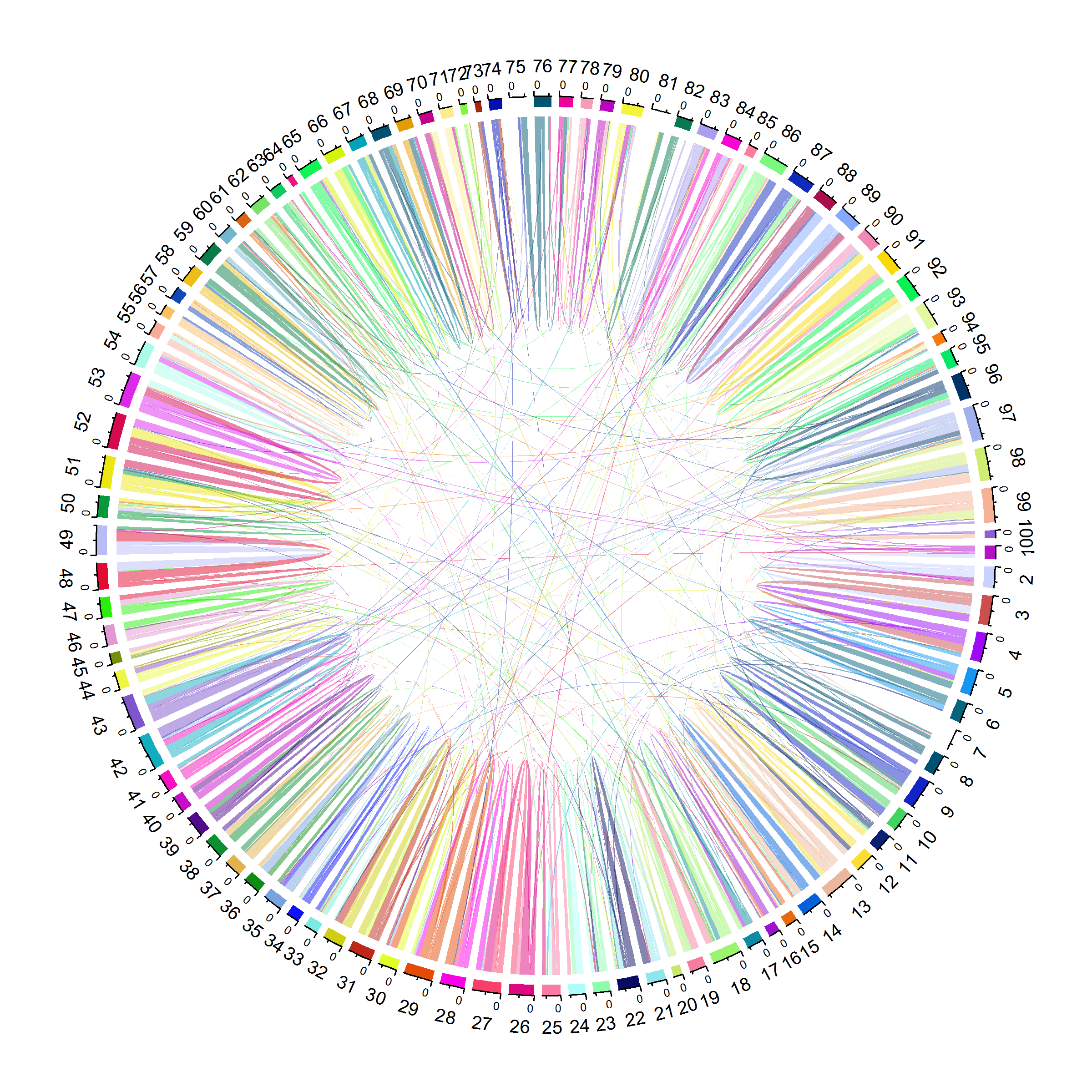}
\caption{nodewise}
\label{fig2.2}
\end{subfigure}
\begin{subfigure}{0.5\linewidth}
\centering
\includegraphics[width = \textwidth]{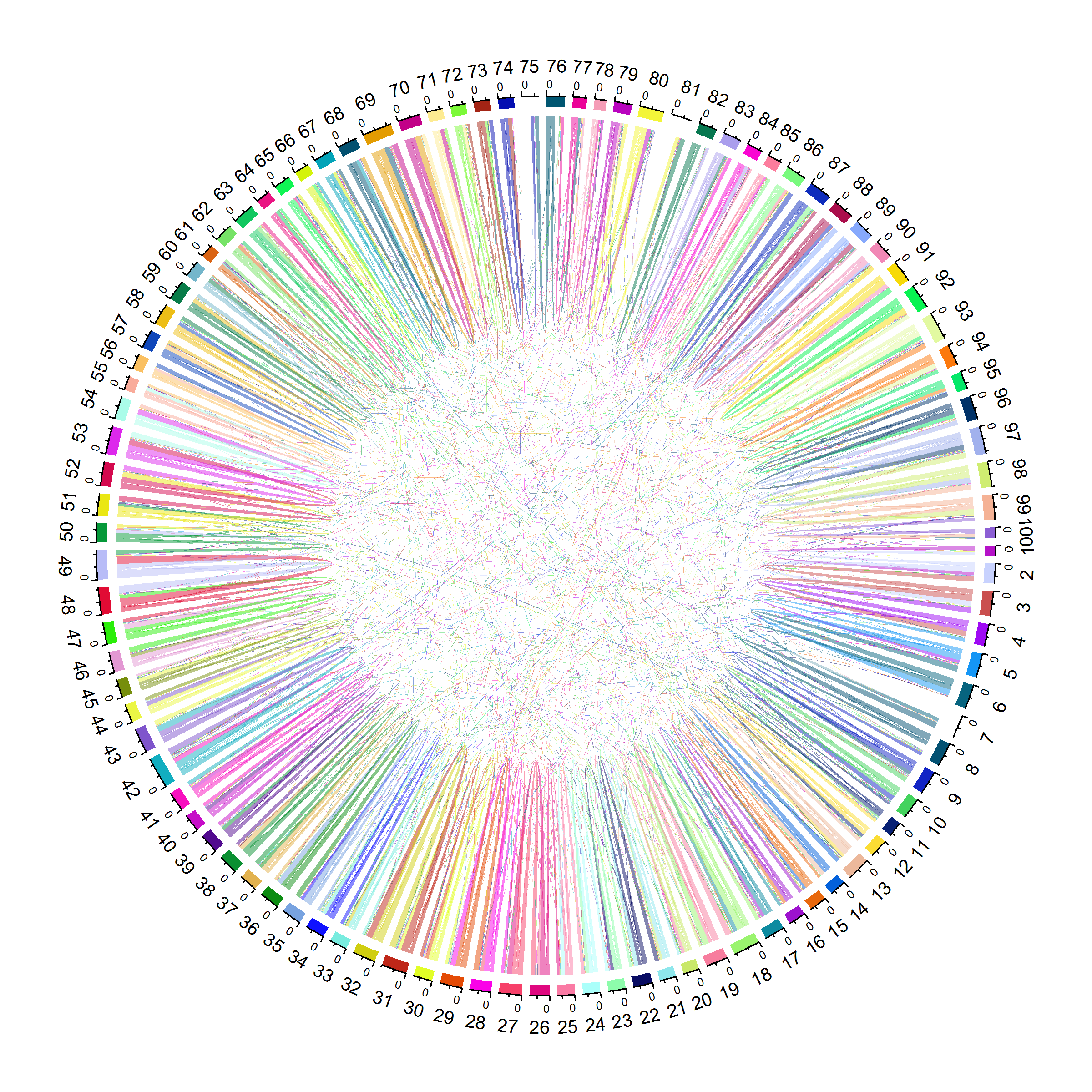}
\caption{space unweighted}
\label{fig2.3}
\end{subfigure}
\caption{Chord diagrams of true and estimated precision matrices corresponding to Toeplitz structure, $n = 200, p = 100$.}
\label{fig2}
\end{figure}

\begin{figure}
\begin{subfigure}{\linewidth}
\centering
\includegraphics[width = 0.5\textwidth]{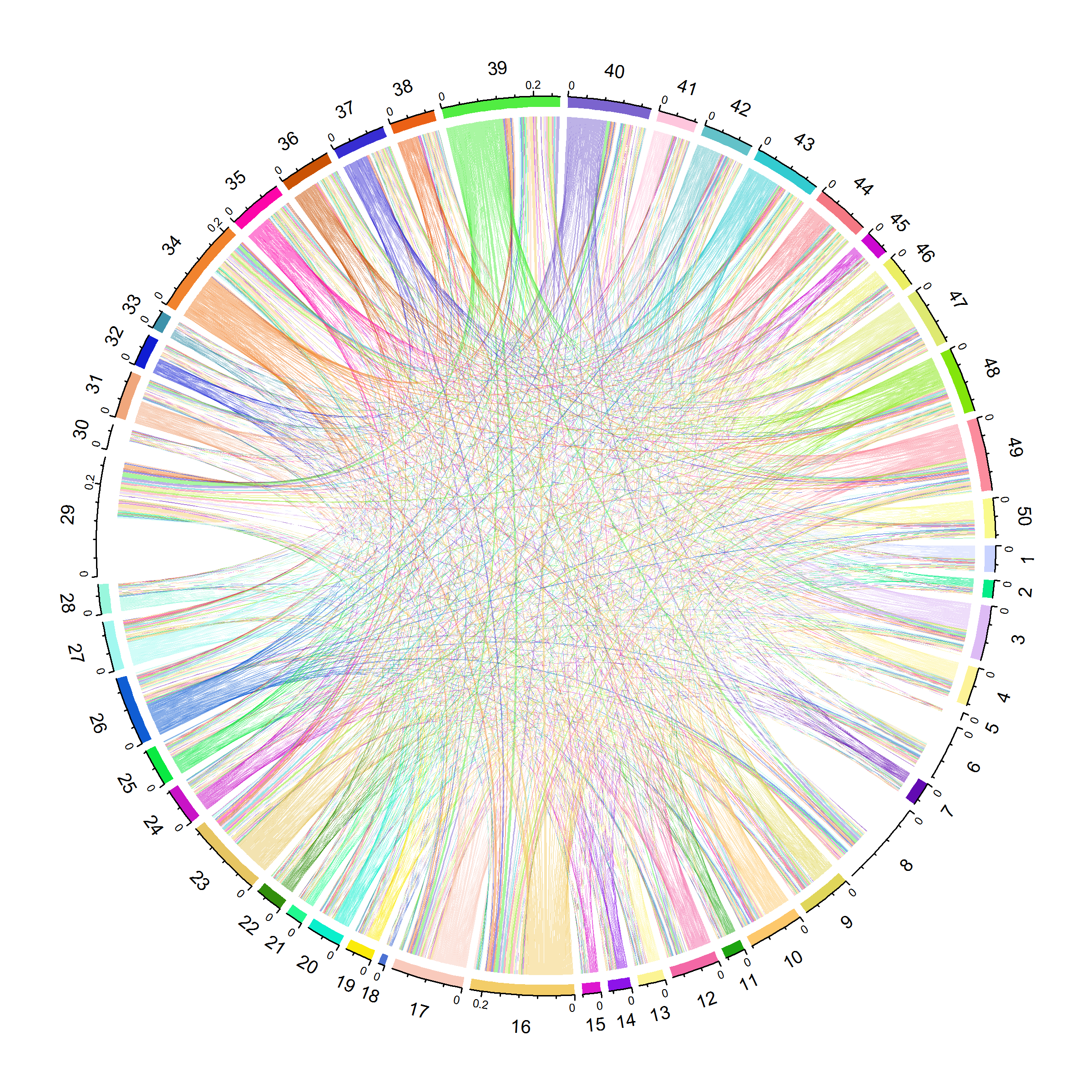}
\caption{true}
\label{fig5.1}
\end{subfigure}\\
\begin{subfigure}{0.5\linewidth}
\centering
\includegraphics[width = \textwidth]{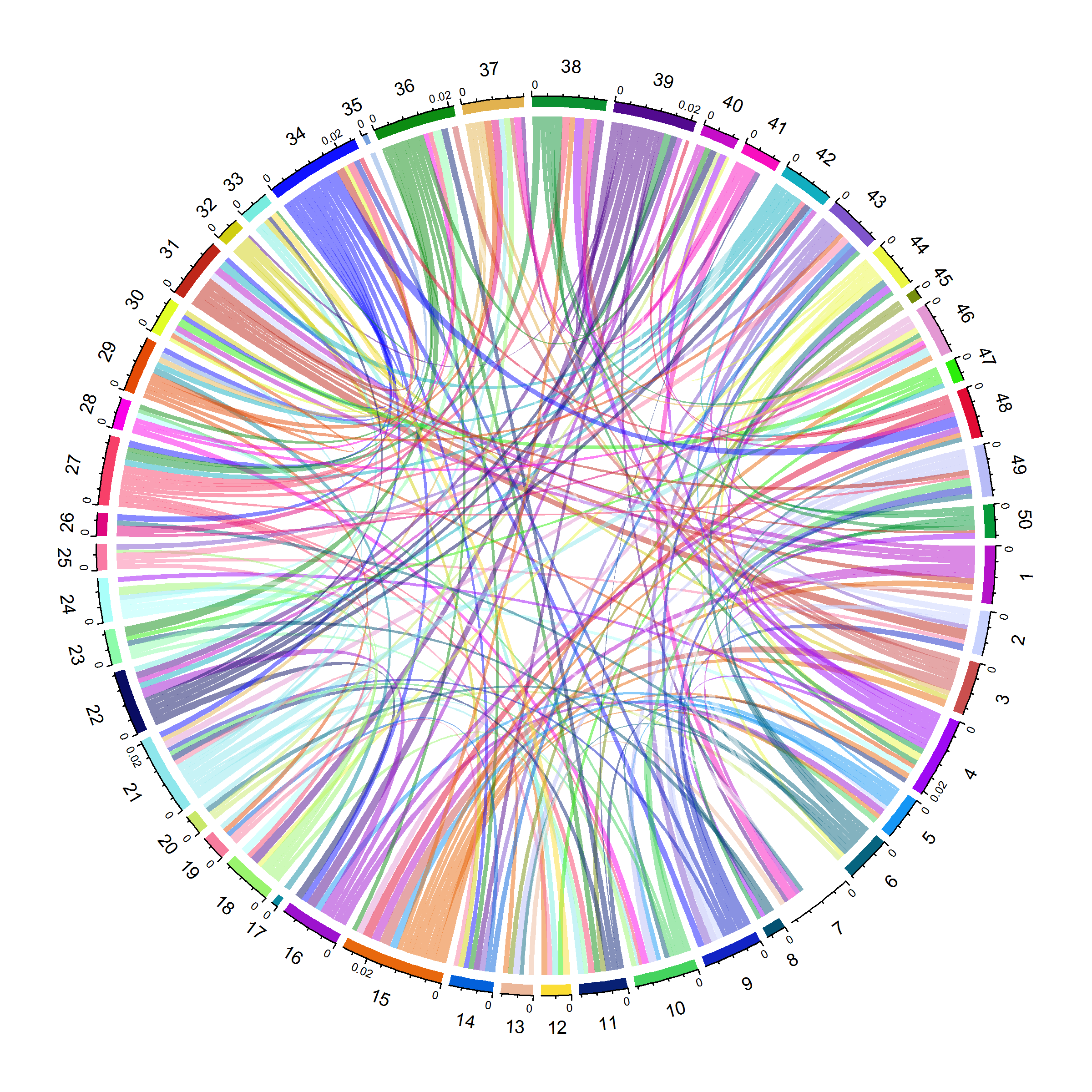}
\caption{nodewise}
\label{fig5.2}
\end{subfigure}
\begin{subfigure}{0.5\linewidth}
\centering
\includegraphics[width = \textwidth]{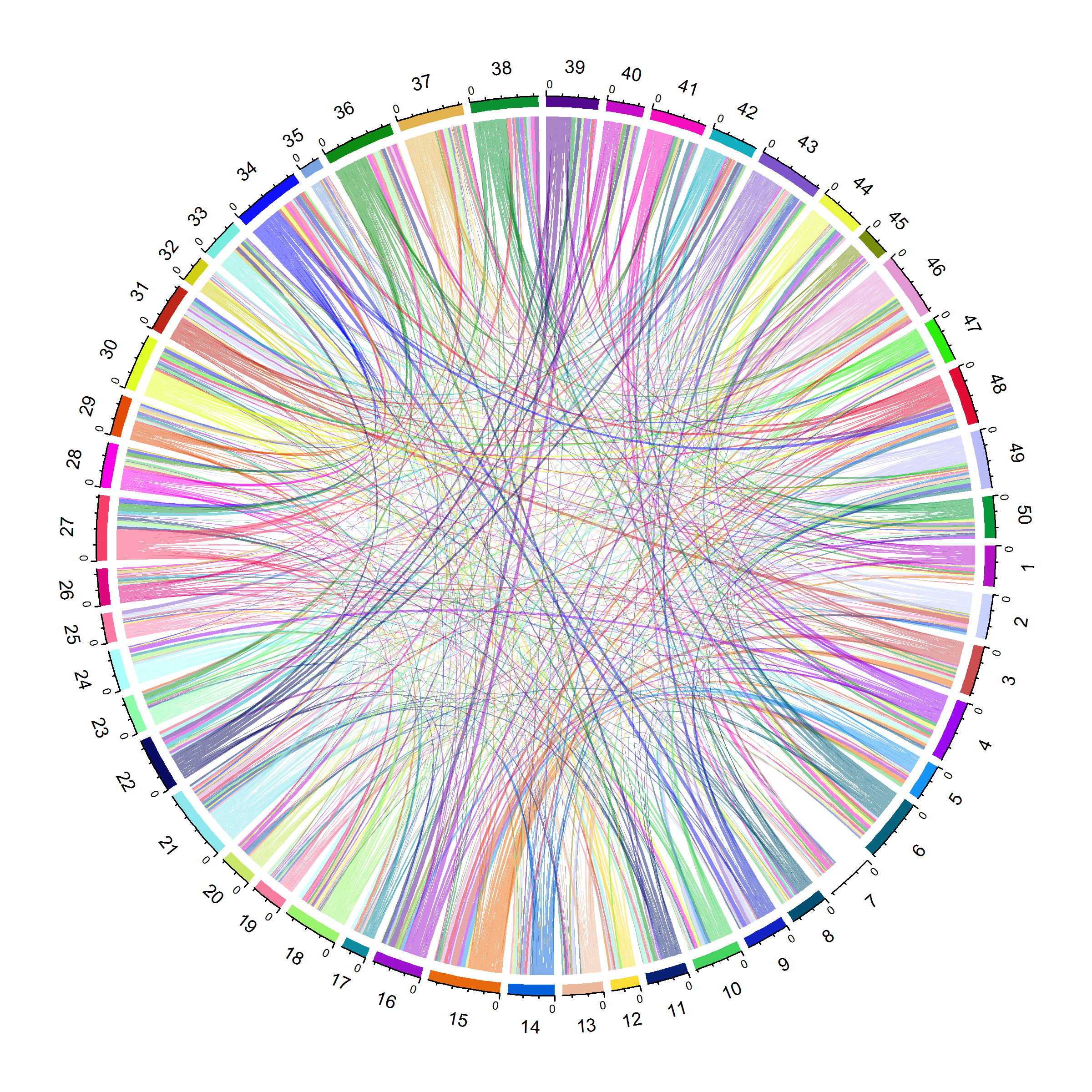}
\caption{space unweighted}
\label{fig5.3}
\end{subfigure}
\caption{Chord diagrams of true and estimated precision matrices corresponding to sparse factor structure of returns, $n = 100, p = 50$.}
\label{fig5}
\end{figure}

\begin{figure}
\begin{subfigure}{\linewidth}
\centering
\includegraphics[width = 0.5\textwidth]{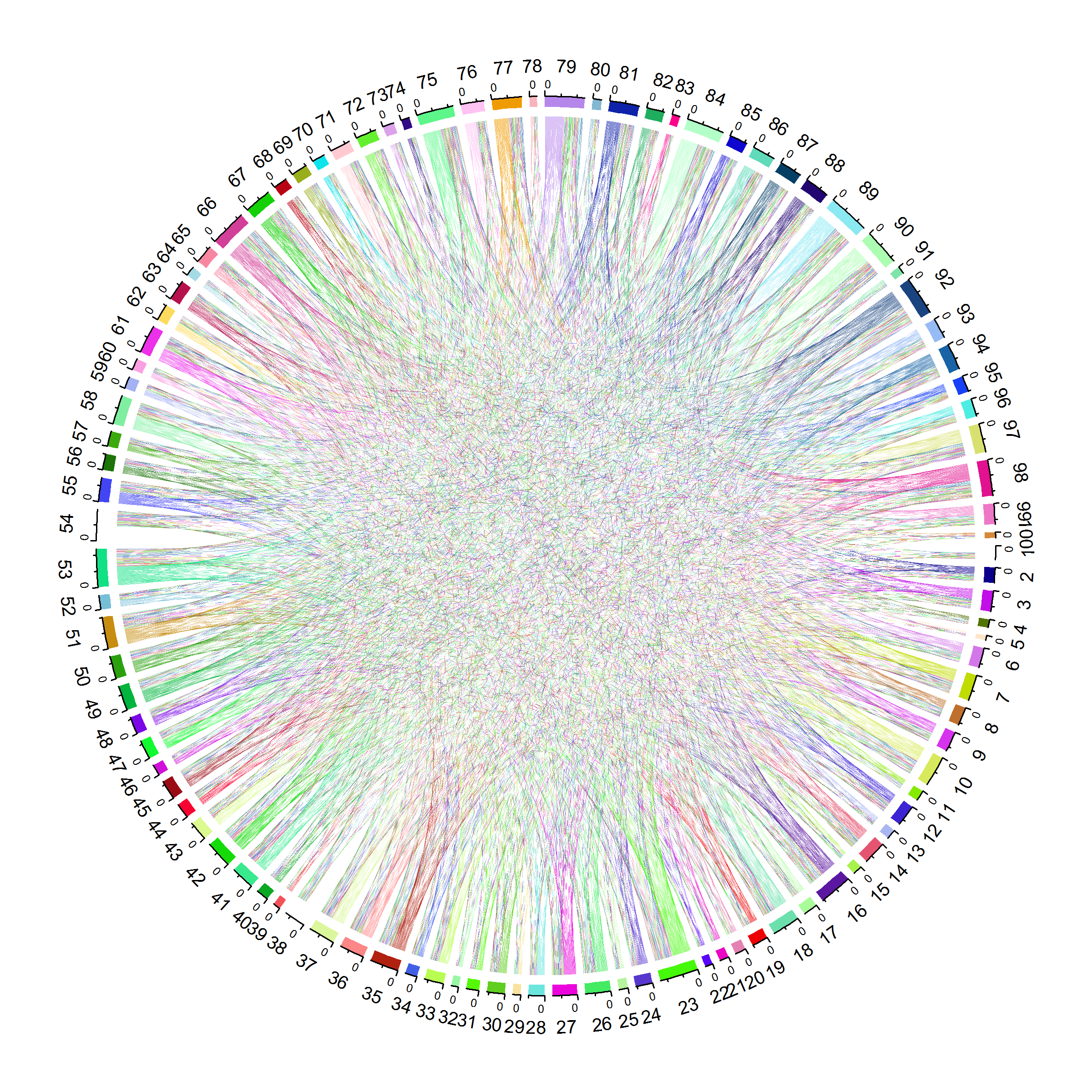}
\caption{true}
\label{fig6.1}
\end{subfigure}\\
\begin{subfigure}{0.5\linewidth}
\centering
\includegraphics[width = \textwidth]{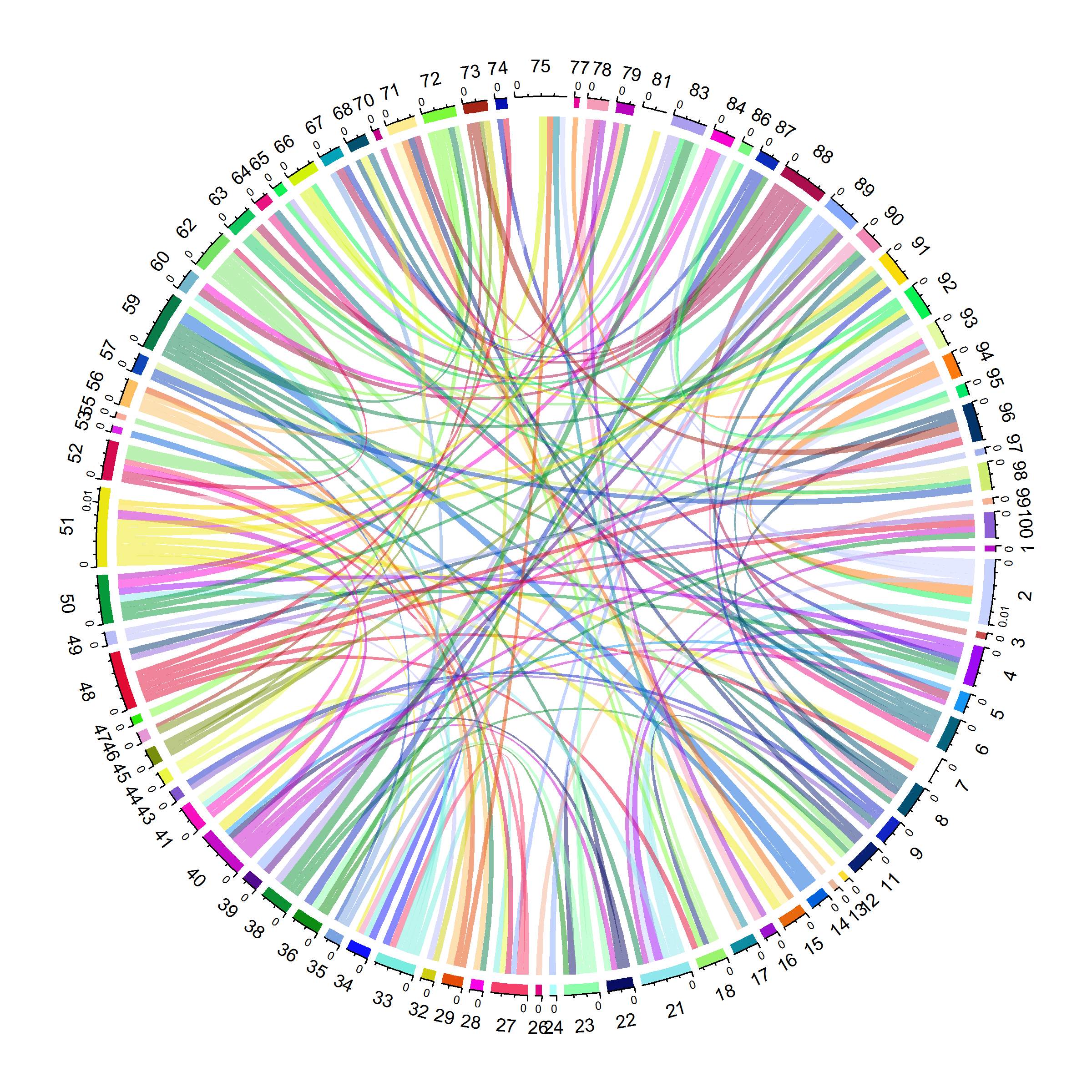}
\caption{nodewise}
\label{fig6.2}
\end{subfigure}
\begin{subfigure}{0.5\linewidth}
\centering
\includegraphics[width = \textwidth]{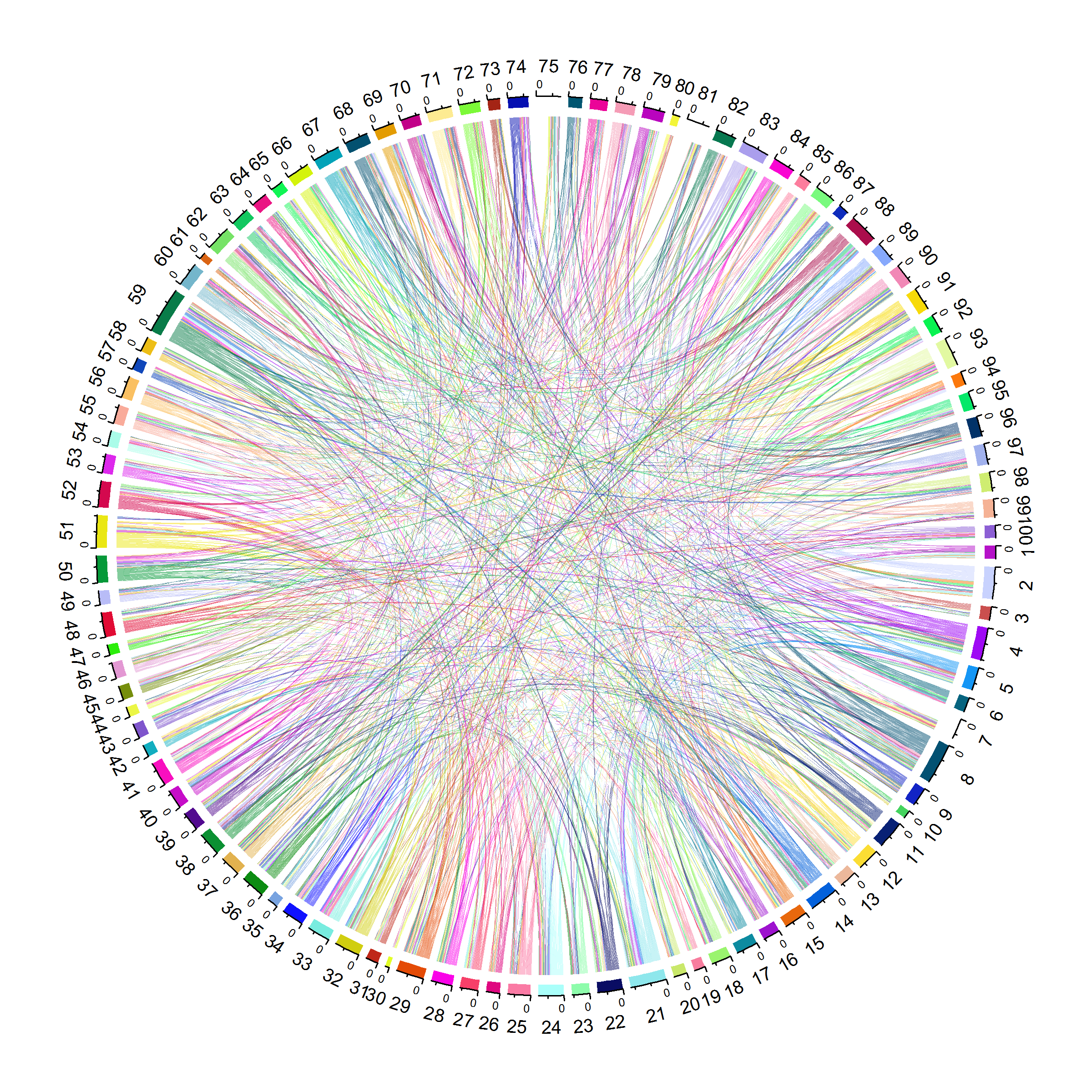}
\caption{space unweighted}
\label{fig6.3}
\end{subfigure}
\caption{Chord diagrams of true and estimated precision matrices corresponding to sparse factor structure of returns, $n = 200, p = 100$.}
\label{fig6}
\end{figure}

In this section, we present visualizations of the estimated precision matrices under the unweighted version of our proposed approach and the nodewise regression method \citep{callot2021nodewise} via chord diagrams. The chord diagrams are presented in Figures~\ref{fig:fig1}, \ref{fig2}, \ref{fig5}, and \ref{fig6}. These visualizations reveal the dependence structure of the underlying assets, and in all the cases we find that the \texttt{space} approach recovers the true precision matrix more accurately as compared to the nodewise regression approach.

\end{document}